# Dynamic frequency dependence of bias activated negative capacitance in semiconductor diodes under high forward bias

Kanika Bansal[1], Mohamed Henini[2], Marzook S. Alshammari[3] and Shouvik Datta[1]

[1]Division of Physics, Indian Institute of Science Education and Research,

Pune 411008, Maharashtra, India

[2]School of Physics and Astronomy, University of Nottingham, Nottingham Nanotechnology and Nanoscience Center, Nottingham NG7 2RD, UK

[3]The National Center of Nanotechnology, KACST, Riyadh 11442, Saudi Arabia

## Abstract

We observed qualitatively dissimilar frequency dependence of negative capacitive response under high charge injection in two sets of junction diodes which are functionally different from each other i.e. electroluminescent diodes and non-luminescent Si-based diodes. Using the technique of bias-activated differential capacitance response, we investigated the mutual dynamics of different rate processes in different diodes. We explain these observations as the mutual competition of fast and slow electronic transition rates albeit differently. This study provides a better understanding of the physics of junction diodes operating under high charge carrier injection and may lead to superior device functionalities.

Behavior of a semiconductor junction diode under high forward bias is something which cannot be considered as a fully understood phenomenon even after decades of research in the field of semiconductors. Junction diodes make the basis of a vast variety of devices which have been an integral part of technology and consumers applications. Some of these junction based devices such as light emitting diodes (LED) and Laser diodes (LD) actually work under high forward bias. Hence it becomes important to understand the electrical impedance of these diodes under high forward bias not only for gaining knowledge about the physical processes they undergo but also for their better use in various applications. In our previous studies[1, 2, 3, 4], we had investigated the impedance of electroluminescent diodes (LEDs and LDs) under charge carrier injection. We observed that as the forward bias increases, the reactive component of the impedance demonstrates negative capacitance. Higher magnitude of this negative capacitance (NC) for lower applied modulation frequencies has also been observed by many groups in electroluminescent devices[5,6,7] and also in other semiconductor devices[8, 9].

Here we report qualitatively opposite dynamical behavior of NC in two sets of diodes with different functionalities: (a) III-V based electroluminescent diodes (ELDs), specifically AlGaInP based multi quantum well lasers (QWL) from Sanyo DL 3148-025 (details can be found in ref. 1) and InGaAs based quantum dot lasers (QDL) grown by Molecular Beam Epitaxy (MBE)[10] and (b) conventional place Si diodes 1N 4001, 1N 4007 and Si based p-i-n photo diode SFH 213, which are intrinsically inefficient for light emission. The modulation frequency dependence of negative capacitance in Si diodes was found to be different than that of ELDs. We extended our earlier understanding[1,2] of negative capacitance as mutual competition of two rate limited processes to explain the observed difference.

Agilent precision LCR Meter E4980A was used for small signal impedance measurements under sinusoidal voltage modulation. The bias value which was applied to the device (as monitored by the LCR meter) is represented as $V_{dc}$.

Figure 1a shows variation of measured small signal capacitance (C) with forward bias for QWL at different modulation frequencies. As the bias increases, the capacitance becomes negative. In log scale, the frequency response curve terminates at a point where it becomes negative. With lower applied modulation frequency, the onset of negative capacitance occurs at lower forward biases. On the other hand, figure 1b shows such capacitance voltage plots for Si diode (1N 4001) under forward bias. Frequency dependence of negative capacitance is completely reversed in this case. As the frequency decreases, we observe negative capacitance onset to shift towards higher applied injection levels. We also measured such negative capacitances in other Si based diodes (p-i-n photo diode SFH 213 and 1N 4007), all of which gave qualitatively similar frequency dependence.

In our earlier studies[1, 2], to explain the occurrence of negative capacitance in light emitting diode structures, we considered the mutual competitive dynamics of sub-band gap defect levels and radiative recombination of available charge reservoir at high injection levels. Depending upon the applied modulation frequency and temperature of the device, defects within a certain energy depth $E_{Th}$ can respond[1] to the applied signal and contribute to the measured impedance. Hence one can write the standard expression relating $E_{Th}$, thermal rate of charge carriers trapping de-trapping from defect states ($\tau$) and modulation frequency ($f$) as:

$$f \approx \frac{1}{\tau} = \nu \exp\left(-\frac{E_{Th}}{k_B T}\right) \qquad (1a)$$

or

$$E_{Th} = k_B T \ln(\nu / f) \qquad (1b)$$

Where $T$ is the temperature which we kept constant (296±0.1K) during the measurements, $k_B$ is the Boltzmann constant and $\nu$ is the thermal prefactor. When the applied modulation is comparable to $1/\tau$, we observe that the defect responds to the applied sinusoidal modulation which ultimately affects the measured impedance of the active junction.

In case of ELDs, the onset of light emission also interrupts the total number of charge carriers available at the junction which includes the contribution from defect levels ($n_{Trapped}$). Radiative recombination process consumes charge carriers coming out of the defect states irreversibly and faster than these are replenished back. As a result, at the end of the modulation cycle, equilibrium defect population is not recovered. This can cause a transient change in the quasi Fermi level (which tracks the charge carrier population) position and in the carrier reservoir at the junction which induces a compensatory lagging behind current to establish the equilibrium. Consequently, we observe an 'inductive like' reactance, experimentally measured as negative capacitance. In addition, the more is the contribution from defect levels ($n_{Trapped}$), the more compensation would be required to achieve such equilibrium. It is easy to understand from equation (1) that with decreasing frequency, the energy depth ($E_{Th}$) increases. This then increases $n_{Trapped}$ and hence the magnitude of compensatory 'inductive like' current which ultimately increases the magnitude of negative capacitance at decreasing modulation frequencies as has been observed in the past[1, 2].

Similar explanation, based on competing different time scale processes, can be extended to understand the impedance response of Si-based diodes too. However, in case of non-luminescent Si-based diodes, we do not have such fast (~ps) radiative recombinations taking place. Still the occurrence of negative capacitance points towards the presence of mutually competing different rate limited electronic processes with different time scales.

To further examine the dynamic dependence of negative capacitance on frequency, in figure 2, we plot and compare the bias values at which the negative capacitance starts to occur ($V_{NC}$) as a function of frequency *f* for two different sets of diodes. In case of ELDs, as shown in figure 2a, $V_{NC}$ increases monotonically with increasing applied frequency. This behavior remains qualitatively the same for quantum well laser diodes (QWL) and quantum dot laser diodes (QDL). This is also in accordance with the work of Feng et al. [11] . It is important to note that the defect contribution can only be activated if they exist below the quasi Fermi level in the band gap[1]. As we increase frequency, we decrease the depth $E_{Th,}$ and hence the contribution from relatively deeper defects is lost. To overcome this, quasi Fermi level has to move towards the band edge to increase the effective defect contribution, which can be achieved by increasing the applied bias (at a constant temperature). Once the defect contribution is significant to compete with the radiative recombination process we start observing NC. This is why we observe that as the modulation frequency increases, $V_{NC}$ increases in ELDs[1].

On the other hand, in figure 2b we observe that for all the Si based 'non-luminescent' devices investigated, $V_{NC}$ does not follow a monotonic behavior with increasing *f*. There, two different regions can be clearly identified depending upon the variation of $V_{NC}$ with frequency. In region I, $V_{NC}$ is relatively independent of applied modulation frequencies for lower frequency ranges below ~$5 \times 10^4$ Hz and remains relatively constant. In region II, above ~$5 \times 10^4$ Hz, $V_{NC}$ decreases with increasing frequencies which is exactly the reverse of what is observed in the case of electroluminescent diodes.

To understand this difference, we studied the frequency derivative of capacitance response. Usually, a peak in *f*dC/d*f* is observed with frequency or temperature variation which corresponds to a position of maximum response from the defects or the position where quasi Fermi level crosses the respective defect energy level[12, 13, 14]. Here in the case of high charge injection, which induces interesting NC behavior, we focus on the bias ($V_{dc}$) activated dynamics of rate processes at room temperature. Experiments showed that[4] this dependences of rate process on $T$ and $V_{dc}$ are mutually inverse and we express it as $1/T=\eta V_{dc}$ where $\eta$ is a proportionality constant for correct dimensionality. Then equation (1) can be modified to the form:

$$f_{Max} \approx \nu \exp\left( -\frac{E_{Th}}{k_B}\eta V_{dc} \right) \quad (2)$$

$f_{Max}$ corresponds to the frequency where maximum response (in *f*dC/d*f*-*f* plot) is observed. Re-writing equation (2) gives us $\ln f_{Max} = \ln \nu + \left( -\frac{E_{Th}}{k_B}\eta V_{dc} \right)$. To verify the validity of this equation one requires a linear behavior of $\ln f_{Max}$ with $V_{dc}$, which we have demonstrated to be true in case of ELDs[4], and therefore the slope ($m$) of the straight line is directly proportional to the thermal activation energy, as given by:

$$Slope(m) = \left( -\frac{E_{Th}}{k_B}\eta \right) \quad (3)$$

Figure 3a shows the variation of *f*dC/d*f* for a wide range of applied modulation frequencies for Si diode 1N 4001. We detect see two prominent peaks in the *f*dC/d*f* vs *f* plot. This is fundamentally different from the differential capacitance response observed in case of ELDs, where only one peak was observed[4]. As the bias increases, $f_{Max}$ for both the peaks shifts to the respective lower frequency sides. This variation of $\ln f_{Max}$ with applied bias is plotted in figure 3b.

Such linear variation of ln $f_{Max}$ with $V_{dc}$ validates our conceived equation 2. Values of the slopes are found to be -4.3±0.1 for lower frequency response and -13.0±0.3 for higher frequency response. This difference may be related to entropic contribution to free energy barrier for a particular electronic defect[15]. Unlike ELDs[4], Si based diode structures do not show any sign change of slopes in ln$f_{Max}$ vs $V_{dc}$ plot which were related to the presence of excitons during light emission. This again emphasizes the point that qualitatively different internal dynamics in ELDs and Si based diodes is responsible for the observed difference in the frequency dependence of negative capacitance response.

It is important to note that in these bias activated processes as the modulation frequency reaches between ~5x$10^4$ and $10^6$ Hz, peak 2 starts to respond significantly only after 0.45 V, however, slower defect response (peak 1) starts appearing at much lower biases. The density of defects responding at such frequencies is directly proportional to the derivative $f$dC/d$f$[12, 14] and hence the area under the peaks corresponds to the density of defects within the energy range spanned by the corresponding range of modulation frequency. In figure 3c, we found that the area acquired by the right half of the peak 1 (which overlaps with peak 2) increases in correlation with the area acquired by the peak 2. Net defect density of peak 1 is much larger as compared to peak 2. However, at large forward biases, the rate of increase in peak 2 is faster than that of peak 1. This can be viewed from the plot in the inset of figure 3c, where values are normalized with respect to the highest area value for respective peaks. We ascribe this observed correlation as an evidence for competition of two rate limited processes governing the voltage onset of respective negative capacitance.

In Si diodes, two differently sized defect responses (bigger peak 1 and smaller peak 2) overlap (within similar frequency/energy ranges) and compete with each other for the overall impedance of the junction. At modulation frequencies > $5x10^4$ Hz, the steady state situation is not recovered at the end of the sinusoidal modulation because the presence of a dominant but slower defect channel interrupts the dynamics of relatively smaller but faster defect response. This results in a mismatch of charge trapping and emission from these two defect channels over a complete cycle. Subsequently, a transient change in the charge carrier population coming from these defects can surface, giving rise to a compensatory lagging behind current and hence the negative capacitance[1, 2] in the measured steady state response of the active junction. In this region II of figure 2b, the contribution from relatively faster defect channels is reduced with decreasing frequency. This requires further increase in the applied bias to generate significant contribution from shallower and faster channels which compete with slower channels to produce any negative capacitance. As a result $V_{NC}$ increases with decreasing frequencies. However, this is qualitatively opposite to the case of ELDs[1] where fast and dominant radiative recombination irreversibly depletes the charge carrier reservoir and competes with much slower and smaller steady state response from defect states.

As we decrease the frequency below ~$5x10^4$ Hz, we modulate the defects which only have slower component of thermal rate of carrier exchange with the band edge. However, until a significant intensity of bias is applied, which can activate the shallower defects/ faster channels we do not observe any NC effect. For high enough forward biases, contribution of faster channels again competes with the slower defect response and NC is observed (figure 2b, region I). Since the activation of this high frequency defect response (peak 2) with increasing bias does not dependent on slower modulation frequencies, we observe that $V_{NC}$ in region I of figure 2b, also does not vary much with changing modulation frequencies. It is important to mention here that in some of the Si-based devices, it has been reported[9, 16] that the frequency dependence of NC is qualitatively similar to that of ELDs. We speculate that this would be due to the interplay of different time scale processes in a manner similar to ELDs.

To summarize, we probed two functionally different diode structures and observed fundamentally different behavior of frequency dependent negative capacitance under high forward biases. We also developed a new approach to probe the device physics by measuring bias activated differential capacitance response ($f$dC/d$f$) as a function of modulation frequency ($f$) which agrees well with the experimental results. In electroluminescent diodes, faster radiative recombinations compete with slower and weaker defect response. In Si diodes, two overlapping defect channels of varied strength and time scales compete and cause negative capacitance. This study provides a better understanding of the electronic processes that give rise to negative capacitance response in junction diodes. These understandings can easily lead to new device functionalities. To further identify the nature of particular electronic defects and thereby optimize the device application, one can extend this study by carefully looking into the necessary details of a typical device structure.

Authors wish to thank IISER-Pune for startup funding of the laboratory infrastructure as well as the grant # SR-S2-CMP-2012 from Dept. of Science and Technology, India. KB is thankful to Council for Scientific and Industrial Research, India for senior research fellowship. MH acknowledges support from the UK Engineering and Physical Sciences Research Council.

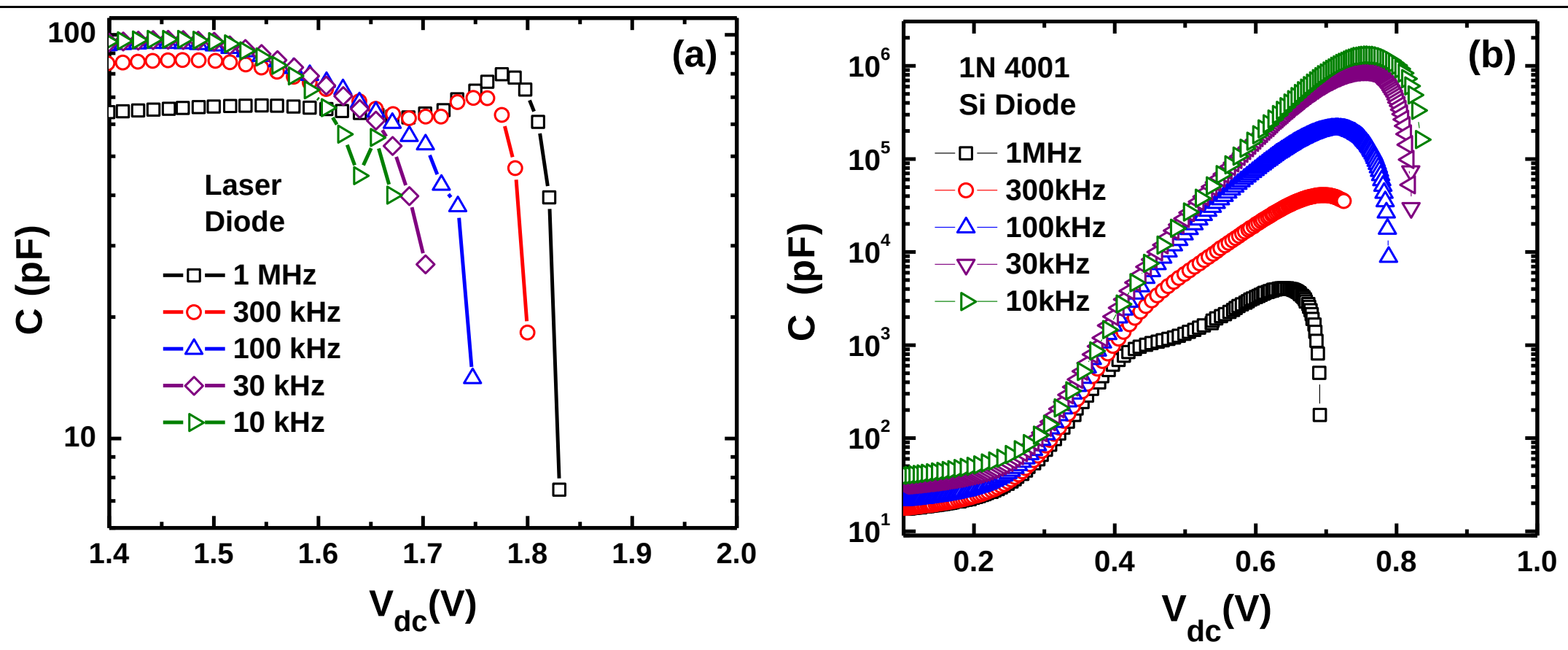

Figure 1: Variation of capacitance with bias under different applied modulation frequencies for (a) quantum well laser diode (DL 3148-025) and (b) Si diode (1N 4001). The point where a particular data plot terminated is the starting of negative capacitance. In Si diodes, high frequency produces negative capacitance for lower biases which is opposite to that of quantum well laser diodes.

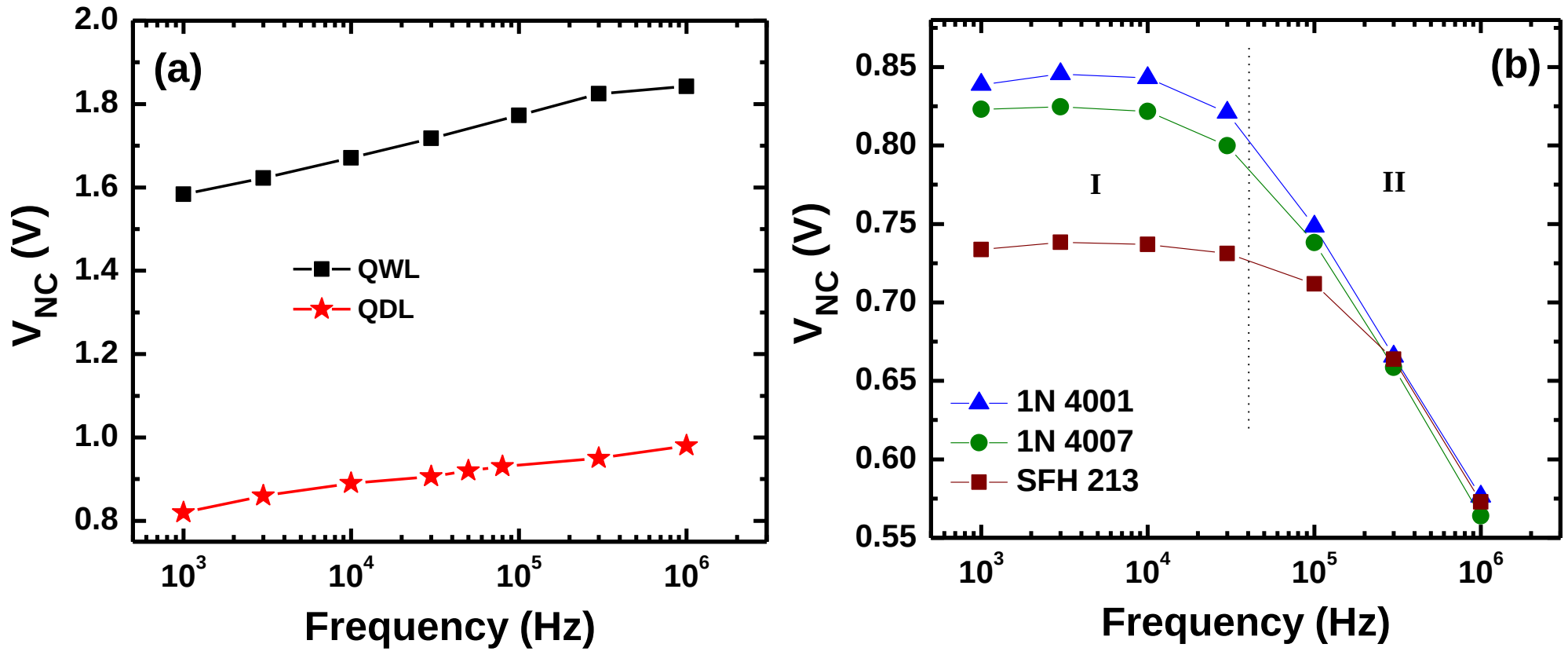

Figure 2: Variation of the bias at which negative capacitance is observed ($V_{NC}$) for different applied modulation frequencies for (a) quantum well laser diodes (QWL) and quantum dot laser diodes (QDL) and (b) Si based diodes. For laser diodes $V_{NC}$ increases monotonically with increasing frequency. For Si diodes $V_{NC}$ first stays somewhat constant with increasing frequency range (region I) and then decreases with increasing frequencies (region II).

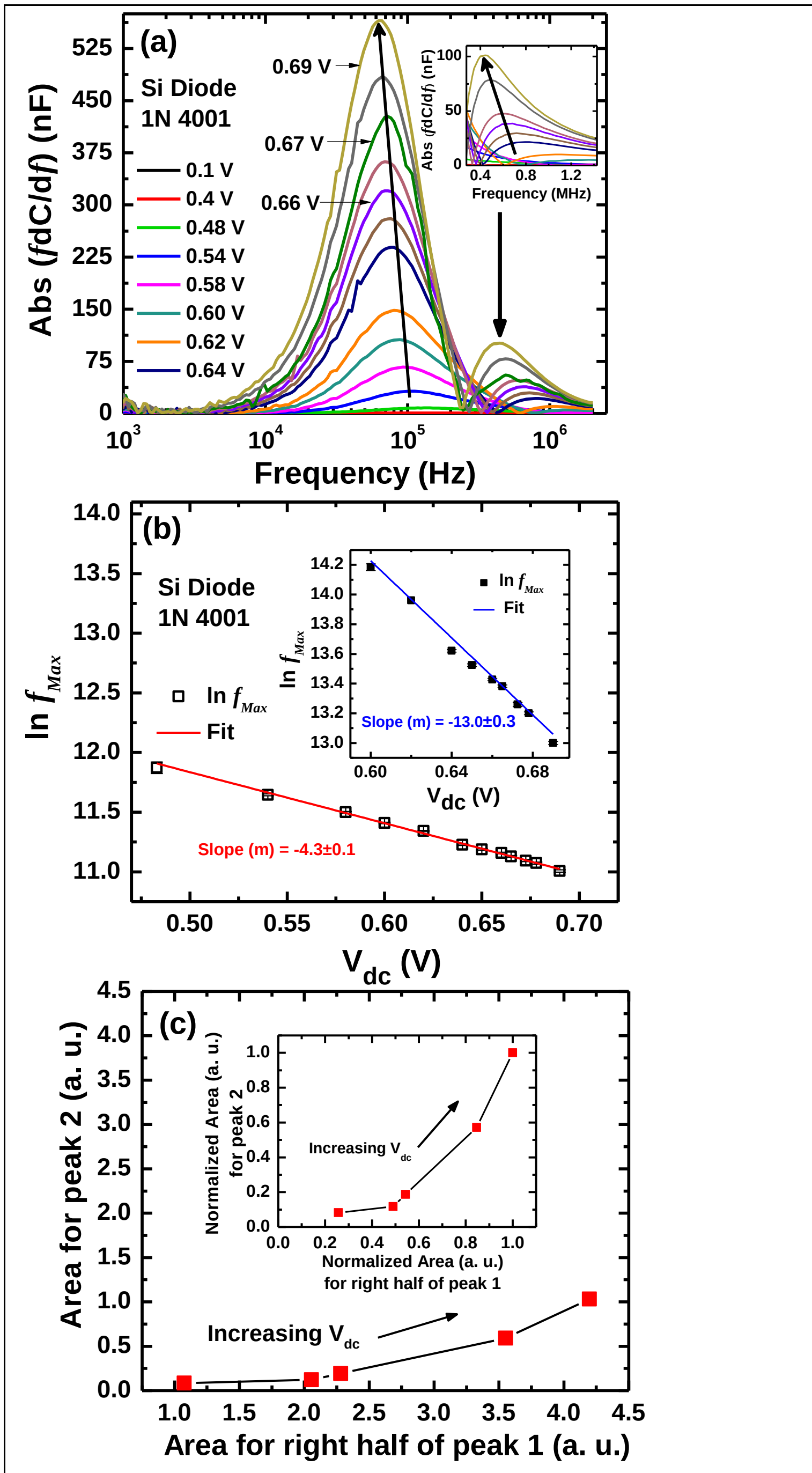

Figure 3: (a) Absolute value of *f*dC/d*f* showing two peaks at different frequency positions for Si diode. Peak frequency values are used to plot ln$f_{Max}$ with applied bias ($V_{dc}$) as shown in (b). Inset in (b) shows the plot for the peaks at higher frequency side (peak 2). In both cases, peak frequency decreases with increasing bias. (c) Shows that the area under the right half of peak 1 and peak 2 increases with increasing bias in a correlated fashion. Lines joining the data points are for visual clarity only.